\documentclass[a4paper]{article}

\newcommand{\preprinttitle}{Population size estimation with capture-recapture in presence of individual misidentification and low recapture}

\newcommand{\listauthors}{\raggedright 
Rémi Fraysse\textsuperscript{1}, \space
Rémi Choquet\textsuperscript{1}, \&
Carlo Costantini\textsuperscript{2}, \&
Roger Pradel\textsuperscript{1}
}

\newcommand{\listinstitutions}{
\textsuperscript{1} CEFE, Univ Montpellier, CNRS, EPHE, IRD -- Montpellier, France
\\
\textsuperscript{2} MIVEGEC, Univ Montpellier, CNRS, IRD -- Montpellier, France
}

\newcommand{\email}{rem.fraysse@gmail.com}

\newcommand{\preprintabstract}{While non-invasive sampling is more and more commonly used in capture-recapture (CR) experiments, it carries a higher risk of misidentifications than direct observations. As a consequence, one must screen the data to retain only the reliable data before applying a classical CR model. This procedure is unacceptable when too few data would remain. Models able to deal with misidentifications have been proposed but are barely used.
Three objectives are pursued in this paper. First, we present the Latent Multinomial Model of Link et al. (2010) where estimates of the model are obtained from a Monte Carlo Markov Chain (MCMC). Second we show the impact of the use of an informative prior over the estimations when the capture rate is low. Finally we extend the model to the multistate paradigm as an example of its flexibility. \\
We showed that, without prior information, with capture rate at 0.2 or lower, parameters of the model are difficult to estimate \textit{i.e.} either the MCMC does not converge or the estimates are biased. In that case, we show that adding an informative prior on the identification probability solves the identifiability problem of the model and allow for convergence. It also allows for good quality estimates of population size, although when the capture rate is 0.1 it underestimates it of about 10\%. 
A similar approach on the multistate extension show good quality estimates of the population size and transition probabilities with a capture rate of 0.3 or more.}

\newcommand{\preprintkeywords}{CMR; environmental DNA; latent multinomial; misidentifications; multistate}

\usepackage[margin=1in,marginparwidth=4.2cm,marginparsep=0.5cm]{geometry}

\usepackage{marginnote}
\usepackage{pdflscape}
\usepackage{ifthen}
\reversemarginpar  
\usepackage{lipsum} 
\usepackage{calc}
\usepackage{siunitx}
\usepackage[right]{lineno}
\usepackage{titlesec}
\usepackage{indentfirst}
\usepackage[utf8]{inputenc}
\usepackage[T1]{fontenc}
\usepackage[
backend=biber,
natbib=true,
sortcites=true,
defernumbers=true,
style=authoryear,
citestyle=authoryear-comp,
maxnames=99,
maxcitenames=2,
uniquename=init,
giveninits=true,
terseinits=true,
url=false
]{biblatex}
\DeclareNameAlias{default}{family-given}
\renewbibmacro{in:}{%
  \ifentrytype{article}{}{\printtext{\bibstring{in}\intitlepunct}}} 
\DeclareFieldFormat[article]{pages}{#1} 
\AtEveryBibitem{\ifentrytype{article}{\clearfield{number}}{}} 
\DeclareFieldFormat[article, inbook, incollection, inproceedings, misc, thesis, unpublished]{title}{#1} 
\usepackage{csquotes}
\RequirePackage[english]{babel} 
\DeclareBibliographyCategory{ignore}
\addtocategory{ignore}{recommendation} 
\usepackage{nameref}
\DeclareFieldFormat{doi}{\url{https://doi.org/#1}}

\usepackage{graphbox}  
\usepackage{microtype}
\DisableLigatures[f]{encoding = *, family = * }
\RequirePackage[default,scale=0.90]{opensans}
\usepackage{xcolor}
\definecolor{darkgray}{HTML}{808080}
\definecolor{mediumgray}{HTML}{6D6E70}
\definecolor{ligthgray}{HTML}{d9d9d9}
\definecolor{pciblue}{HTML}{74adca}
\definecolor{opengreen}{HTML}{77933c}
\usepackage{changepage}
\usepackage[aboveskip=1pt,labelfont=bf,labelsep=period,singlelinecheck=off]{caption}
\usepackage{amsmath}
\usepackage{amsfonts}

\usepackage[pdfborder={0 0 0},   
    colorlinks=true,
    linkcolor=blue,
    urlcolor=blue,
    citecolor=blue]{hyperref}  

\newcommand{\beginingpreprint}{
\vspace*{0.5cm}
\begin{flushleft}
\baselineskip=0pt

{\Huge
\fontseries{sb}\selectfont{\preprinttitle}}
\end{flushleft}
\vspace*{0.25cm}
\begin{flushleft}
\Large
\listauthors
\end{flushleft}
\bigskip
{\raggedright
\listinstitutions}
\\
\\
\textbf{Correspondence: } \href{mailto:\email}{\email}\\
\\
\vspace*{0.5cm}
\fcolorbox{ligthgray}{ligthgray}{
\parbox{\textwidth - 2\fboxsep}{
\vspace{0.25cm}
\begin{internallinenumbers}
\textbf{\large{\textsc{Abstract}}}\\

\preprintabstract\\

\textbf{\emph{Keywords: }}\preprintkeywords
\end{internallinenumbers}

\vspace{0.25cm}}
}
\newpage
\newgeometry{margin=1in}
}

\begin{document}
\beginingpreprint



\section*{\centering Introduction}

Individual identification based on natural tags is widely used in capture-recapture studies, either for estimating population size or survival. Natural tags can be environmental DNA (eDNA) - examples of such studies are on bears \parencite{dreher_noninvasive_2007}, bobcats \parencite{ruell_estimating_2009, morin_efficient_2018}, pronghorns \parencite{woodruff_estimating_2016}, and elephants \parencite{laguardia_nationwide_2021} - 
or visual patterns - examples of such studies are on whales \parencite{curtis_abundance_2021}, dolphins \parencite{labach_distribution_2022}, leopards \parencite{swanepoel_density_2015} and beetles \parencite{quinby_estimating_2021}.
Although non-invasive sampling allows studying free-ranging, elusive species without having to catch, handle or even observe them, there are still some difficulties to be confronted to. In particular, compared to traditional tagging methods, there is a much higher risk of incorrect individual identification when tags are based on natural features \parencite{taberlet_noninvasive_1999}. If misidentifications are ignored, classical models overestimate population size, up to five fold \parencite{creel_population_2003}. 
For eDNA sampling, several studies have proposed solutions to reduce misidentification, from field methods and good laboratory techniques for genetic analyses \parencite{paetkau_empirical_2003, waits_noninvasive_2005} to pre-analysis software that help filter out data that are likely to contain errors \parencite{mckelvey_dropout_2005}. Regarding visual patterns recognition, computer-aided image matching processes \parencite{bolger_computer-assisted_2012, crall_hotspotter_2013} have been developed to help with the identification, and an analysis R package have been developed to deal with data where photos from left and right side of the individuals are available without a reliable mean to match them \parencite{mcclintock_multimark_2015}.
In addition, various proposals have been made to account for misidentifications in models estimating population size \parencites{lukacs_estimating_2005, wright_incorporating_2009, link_uncovering_2010, yoshizaki_modeling_2011}.
Today, the most common practice remains the filtering out of low quality photo or eDNA samples that were not sufficiently amplified. \\

Depending on the percentage of low quality data, discarding may lead to retaining too little data for a reliable estimation of the parameters of interest. 
Of the studies using natural tags cited previously, five estimated recapture rates under 0.2 while rejecting between 20 and 40\% of the collected samples. Their low catch rate can be explained by two factors: the large populations (over 1800 individuals to more than 70,000 in the case of \cite{laguardia_nationwide_2021}) or the elusive species (like whales). 
In such cases, it may be beneficial to allow a small degree of uncertainty in the identification as proposed by \cite{lukacs_estimating_2005}, around 1-5
It is possible to model this error rate. If the cost of adding a parameter (the error rate) is offset by the number of samples it allows to keep, then the trade-off is interesting. 
When studying a large population or having a complex observation process, keeping more samples with lower quality could be necessary while still not leading to many recaptures. In such a situation, it is necessary to know how the chosen model performs. Among the approaches that incorporate the misidentification process into the analysis model, \cite{yoshizaki_modeling_2011} already suggests that their least square method does not perform well with few recapture. Wright's model \parencite{wright_incorporating_2009} requires genotypes replicates to estimate an error rate which increases costs, especially for large populations where a lot of samples are obtained. 
On the other hand, the Latent Multinomial Model (LMM, \cite{link_uncovering_2010}) is a malleable framework that has received attention and has been extended by other publications \parencite{mcclintock_probit_2014, schofield_connecting_2015, bonner_extending_2015}. It estimates, in a Bayesian framework, the misidentification rate without additional information. For its flexibility, its way of estimating parameters and the attention it has received, this model seems promising. However we do not know how it performs with few recaptures.  \\

This study is part of a project aiming to apply capture recapture to mosquito larvae using eDNA. For this project we expect having very few eDNA per sample, which would imply having to discard a large proportion of them.
Augmenting the effective number of captures can be achieved by augmenting the capture effort or by keeping more samples by keeping lower quality ones. The first approach imply increasing the financial cost. Moreover, the logistic associated with collecting more samples that need to be dealt with in a proper timing or stored appropriately could be prohibitive on some fields like in developing countries. Thus we focus on the second approach and model the identification errors in order to use as many samples as possible. 
However, even when keeping lower quality samples, we are still expecting low numbers of recapture.
To design an experiment with few recaptures, one needs to know the capabilities of the model in expected ranges of capture and identification probabilities. As such, this paper gives, through the use of simulations, some guides and limits to use the LMM when confronted to low capture probabilities as well as a possible solution to complement the lack of information in the data when it happens, through the use of an informative prior. We also extend the model to multistate observations to show how the model framework can be used in more complex cases but more general biological situations. \\

We start by describing the multinomial model $M_t$ that estimates population size with time-varying detection rates, and the latent multinomial model $M_{t,\alpha}$ that deals with misidentifications. We then extend the model $M_{t,\alpha}$ to multistate observations. Finally, we present the simulations we made and how the model performs estimating the population size, with and without informative priors. \\

\section*{\centering Closed single state models}

\subsection*{\centering Model $\mathbf{M_t}$}

When estimating population size $N$ in a closed capture-recapture experiment (the population is assumed not to change), with the model called $M_t$ \parencite{darroch_multiple-recapture_1958, otis_statistical_1978}, individuals are assumed to be observed ("captured") with probability $p_t$ at occasion \emph{t} for $t=1,2,...,T$ and identified individually through human made or natural marking. Capture events are supposed independent among individuals and over time. \\

For each occasion, individuals are assigned a 0 if they were not captured or a 1 if they were. This leads to $2^T$ possible distinct histories, including the non observable all-zero one. They are represented by the sequences $\boldsymbol\omega_i=(\omega_{i,1}, ...,\omega_{i,T})$. We reference the observable histories through their index $i = \sum_{t=1}^T \omega_{i,t} \cdot 2^{t-1}$.
Let $y_i$ be the number of individuals with history $\omega_i$ and $\mathbf{y} = (y_1, y_2,...,y_{2^T-1})$. \textbf{y} follows a multinomial distribution with index $N$ and cell probabilities
\begin{equation}\label{eq:pHistMt}
\pi_i = \prod_{t=1}^T \left[ p_t^{I(\omega_{i,t}=1)}  (1-p_t)^{I(\omega_{i,t}=0)} \right]
\end{equation}
where $I(test)$ is 1 if \emph{test} is true, 0 otherwise.

\subsection*{\centering Model $\mathbf{M_{t,\boldsymbol\alpha}}$}\label{sec:Mta}

To account for individual misidentifications, \parencite{yoshizaki_modeling_2011} proposed a model $M_{t,\alpha}$ where captured individuals are correctly identified with probability $\alpha$. Misidentifications are assumed to always create a new individual (a "ghost"). An individual cannot be mistaken as another and two errors cannot create the same ghost. 
To estimate the parameters of the model, \parencite{link_uncovering_2010} developed a latent structure to the model $M_{t,\alpha}$, allowing for a bayesian estimation of the parameters. In this structure, misidentifications are denoted by 2's in latent error histories. These latent error histories $\boldsymbol\nu_j = (\nu_{j,1}, ..., \nu_{j,T})$ are referenced by index $j = 1 + \sum_{t=1}^T \nu_{j,t} \cdot 3^{t-1}$ and $x_j$ ($\mathbf{x} = (x_1, ..., x_{3^T})$) is the number of individuals with latent error history $\nu_j$.
In order to break down the likelihood into two parts, the capture process and the identification one, and to make future developments of the model easier, we follow \parencite{bonner_extending_2015} by introducing latent capture histories $\boldsymbol\xi_k = (\xi_{k,1},...,\xi_{k,T})$. They are the true capture histories, \emph{i.e.} in absence of individual misidentifications, composed of 0 and 1. They are referenced by index $k=1+\sum_{t=1}^T \xi_{k,t} \cdot 2^{t-1}$ and $z_k$ ($\mathbf{z} = (z_1, ..., z_{2^T})$) is the number of individuals with latent capture history $\xi_k$. \\

In this model framework, the observed frequencies \textbf{y} are a known linear transformation \textbf{y}=\textbf{Ax} of the latent error histories frequencies \textbf{x} for a given matrix \textbf{A}. The constraint matrix \textbf{A} is $(2^T-1) \times 3^T$ with a 1 at row \emph{i} and column \emph{j} if the latent error history \emph{j} gives rise to the observed one \emph{i}. All the other entries are zeros.
The latent capture frequencies \textbf{z} are another linear transformation \textbf{z}=\textbf{Bx} for a given matrix \textbf{B}.
 \textbf{B} is $2^T \times 3^T$ with 1 at row \emph{k} and column \emph{j} if the latent capture history $\xi_k$ and the latent error history $\nu_j$ have the same capture pattern. \\

The conditional likelihood is
\begin{equation}\label{eq:Likelihood}
	[\mathbf{y} | \mathbf{x}, \mathbf{z}, N, p, \alpha] = 
	  I(\mathbf{y=Ax}) \, [\mathbf{x} | \mathbf{z}, \alpha] \, [\mathbf{z}|N, \mathbf{p}]
\end{equation}

The capture process is the same as in model $M_t$, using histories $\xi$ and frequencies \textbf{z}. The capture likelihood is the following multinomial product where $\pi_k$ are computed as in \ref{eq:pHistMt}, using histories $\xi$ instead of $\omega$:

\begin{equation}\label{eq:Likelihoodz}
[\mathbf{z} | N, \mathbf{p}] = \frac{N!}{\prod_k z_k!} \prod_{k} \pi_k^{z_k}
\end{equation}

\cite{bonner_extending_2015} gives the likelihood of the identification process, knowing the real captures:

\begin{equation}\label{eq:likelihoodx}
	[\mathbf{x} | \mathbf{z}, \alpha] = I(\mathbf{z}=\mathbf{Bx})
	  \frac{\prod_k z_k!}{\prod_j x_j!} 
	  \prod_{j}\left[ \prod_{t=1}^T A_{j,t} \right]^{x_j}
\end{equation}

with $A_{j,t} =  \alpha^{I(\nu_{j,t}=1)}  (1-\alpha)^{I(\nu_{j,t}=2)}$. The ratio of factorials accounts for the many relabellings of the marked individuals that would produce the same counts in \textbf{x} and \textbf{z}. \\

The full likelihood is obtained by summing the conditional one $[\mathbf{y} | \mathbf{x}, \mathbf{z}, N, p, \alpha]$ over all values of \textbf{x} belonging to the set $\mathcal{F}_{\mathbf{y}} = \{\mathbf{x}|\mathbf{y}=\mathbf{Ax}\}$:
\begin{equation}\label{eq:fullLikelihood}
	[\mathbf{y} | N, p, \alpha] = \sum_{x \in \mathcal{F}_{\mathbf{y}}} [\mathbf{y} | \mathbf{x}, \mathbf{z}, N, p, \alpha]
\end{equation}

\subsection*{\centering Bayesian estimation of the parameters}\label{sec:mcmc}

The feasible set $\mathcal{F}_{\mathbf{y}}$ is complicated to enumerate, which makes the likelihood (eq. \ref{eq:fullLikelihood}) almost untractable in terms of computation. MLE is thus not practical. Conveniently, \cite{link_uncovering_2010} show how a Markov Chain Monte Carlo (MCMC) can be constructed in a bayesian analysis. The Markov chain will allow for the estimation of the posterior density:
\begin{equation}
	[N, \mathbf{p}, \alpha | \mathbf{y}] \propto 
	  [\mathbf{y} | N, \mathbf{p}, \alpha] \, [N] \, [\mathbf{p}] \, [\alpha],
\end{equation}

where $[N]$, $[\mathbf{p}]$ and $[\alpha]$ denote the priors on population size, capture probability and identification probability. \\

Let $\beta(a_0^t, b_0^t)$ denote the beta prior on $p_t$ and $\beta(a_0^\alpha, b_0^\alpha)$ denote the beta prior on $\alpha$. As showed by \cite{link_uncovering_2010}, the likelihood being multinomial, it follows that these priors lead to full conditional distributions
$p_t \sim \beta(a_0^t+a^t, b_0^t+b^t)$ where
$a^t$ is the number of captured individuals at time $t$ and $b^t$ the number of unseen individuals at time $t$ (including the individuals never seen), and 
$\alpha \sim \beta(a_0^\alpha + a^\alpha, b_0^\alpha + b^\alpha)$ where
$a^\alpha$ is the total number of correct identifications  and
$b^\alpha$ the total number of misidentifications. Thus,
\begin{itemize}
	\item $a^t = \sum_k z_k I(\xi_{k,t}=1)$
	\item $b^t = \sum_k z_k I(\xi_{k,t}=0)$
	\item $a^\alpha = \sum_j x_j I(\nu_{j,t}=1)$
	\item $b^\alpha = \sum_j x_j I(\nu_{j,t}=2)$
\end{itemize}

\emph{N} has to be sampled jointly with \textbf{x} since the number of errors in \textbf{x} changes N. Sampling \textbf{x} requires to be able to sample from $\mathcal{F}_{\mathbf{y}}$. \cite{link_uncovering_2010} proposed sampling moves from the null space of matrix \textbf{A}, 
\[
Ker_{\mathbb{Z}}(\mathbf{A}) = Ker(\mathbf{A}) \cap \mathbb{Z}^d 
                             = \{\mathbf{x} \in \mathbb{Z}^d|\mathbf{Ax}=0\},
\]
and adding or substracting them to the current \textbf{x} in the MCMC.
\cite{schofield_connecting_2015} showed that if the basis of $Ker_{\mathbb{Z}}(\mathbf{A})$ was not carefully selected, some parts of the space $\mathcal{F}_{\mathbf{y}}$ could be disconnected from the others and the Markov chain would only explore sub-spaces, depending on the initial \textbf{x}, possibly leading to biased estimations. They proposed to sample moves from the Markov basis of \textbf{A} \parencite{diaconis_algebraic_1998}, a set in $Ker_{\mathbb{Z}}(A)$ that connect all $\mathcal{F}_{\mathbf{y}}$ irrespective of the values in \textbf{y}. Such a basis ensure that the whole set $\mathcal{F}_{\mathbf{y}}$ is connected by single moves and that no move will get out of the set. The drawback is that the computation of that markov basis is heavy and algebraic softwares such as 4ti2 \parencite{4ti2} will not be able to calculate it for $T \ge 5$. \\

\cite{bonner_extending_2015} proposed a mechanism to avoid computing that basis. It consists in sampling from dynamic Markov basis \parencite{dobra_dynamic_2012} which is the set of moves $M(x)$ that connect each \textbf{x} to some neighbours. The algorithm is randomly adding or removing an error from the set of latent histories. To add an error, the authors choose a history that may have generated a ghost (\emph{i.e.} a history containing a 0), and "merge" it with a potential ghost (\emph{i.e.} replace the 0 by a 2 and remove the ghost history). To remove an error, they choose a history containing a 2, replace it by a 0 and add a history with a unique capture (coded 1) at that time. \\

Formally, let's define $\nu_{1t}$ the history with a unique capture at time $t$ (potential ghost), $X_{0,t}(\mathbf{x})=\{\nu |\nu_t=0, x_{\nu}>0, x_{\nu_{1t}}>0 \}$ the set of histories having \emph{potentially} generated a ghost at time \emph{t}, for the given \textbf{x} and $X_{2,t}(\mathbf{x})=\{\nu |\nu_t=2, x_{\nu}>0\}$ the set of histories \emph{containing} a ghost at time \emph{t}, for the given \textbf{x}.
The mechanism to add an error is:
\begin{itemize}
	\item Sample $\nu_0 \in X_{0.}(x) = \bigcup_t X_{0,t}(x)$.
	\item Sample $t \in \{t| \nu_{0,t}=0, x_{\nu_{1t}}>0\}$.
	\item Define $\nu_2 = \nu_0+2\nu_{1t}$.
	\item Define the move $b_{\nu_0, \nu_1, \nu_2}=(-1,-1,+1)$.
\end{itemize}

The mechanism to remove an error is:
\begin{itemize}
	\item Sample $\nu_2 \in X_{2.}(x) = \bigcup_t X_{2,t}(x)$.
	\item Sample $t \in \{t| \nu_{2,t}=2\}$.
	\item Define $\nu_0 = \nu_2-2\nu_{1t}$.
	\item Define the move $b_{\nu_0, \nu_1, \nu_2} = (+1,+1,-1)$.
\end{itemize}

The proposal vector of latent histories $\mathbf{x}'$ is defined as $\mathbf{x}^{(k-1)} + b$ and $\mathbf{z}'$ is calculated with $\mathbf{z}'=\mathbf{Bx}'$.
The $\mathbf{x}'$ and $\mathbf{z}'$ are then accepted or rejected through the Metropolis-Hastings algorithm with probability

\begin{equation}\label{eq:MHratio}
	r_1 = min\left(1, 
	  	\frac{[\mathbf{y} | \mathbf{x}', \mathbf{z}', N', p, \alpha]}
	       {[\mathbf{y} | \mathbf{x}^{(k-1)}, \mathbf{z}^{(k-1)}, N, p, \alpha]}
	  \frac{q(\mathbf{x}^{(k-1)}|\mathbf{x}')}
	       {q(\mathbf{x}'|\mathbf{x}^{k-1})}
	\right)
\end{equation}

The proposal densities are calculated by multiplying the probabilities of each sampling step used for defining the move. They are successively: the probability of adding (or removing) an error, the probability of choosing the $\nu_0$ (or $\nu_2$) and the probability of choosing the \emph{t} knowing the sampled $\nu$. When adding an error, the proposal density $q$ is:
\begin{equation}\label{eq:q-add}
	q(x^{prop}|x^{k-1}) = 
	    \frac{0.5}
	         {\#X_{0.} \#\{t|\nu_{0,t}=0,x_{\nu_{1t}}>0\}}
\end{equation}
and when removing an error, is:
\begin{equation}\label{eq:q-rem}
q(x^{prop}|x^{k-1})=\frac{0.5}{\#X_{2.} \#\{t|\nu_{2,t}=2\}}
\end{equation}
where $\#S$ denotes the cardinality of $S$.\\

We need to add a last Metropolis-Hastings sampler to sample the number of unseen individuals. A move $c$ can be sampled in $[-D, D]$ where D is a fixed hyperparameter and defining $n_0' = n_0 +c$. If $n_0' \ge 0$, it is accepted with probability $r_2$ which is defined as $r_1$ in equation \ref{eq:MHratio}. Since only the number of unseen individuals changes and that the proposal density is symmetric, $r_2$ simplifies as:
\begin{equation}
	r_2 = min\left(1, 
	  \frac{[ \mathbf{z}'  | N', p ]}
	       {[ \mathbf{z}  | N, p ]}
	\right).
\end{equation}

\section*{\centering Closed multistate models}
\subsection*{\centering Arnason-Schwarz model}

The time-dependent multistate Arnason-Schwarz model assumes individuals to move independently over a finite set of S states, $E=\{e_1,...e_S\}$. These states are not observed at each occasion for every individual but only when they are captured. Capture histories $\omega_i$ are now composed of \emph{S+1} values. The \emph{1, ..., S} are used when the individuals are seen in states $e_1,...e_S$ and the 0 when the individuals are not seen. We assume that the state is always correctly identified on capture. We now have $p_{s,t}$, the detection probabilities that vary both in time (denoted as before \emph{t}) and in states (denoted s). We note
\begin{itemize}
	\item $\psi_{s,r}$ the probability of being in state $e_r$ at time $t+1$ if in state $e_s$ at time $t$ (\emph{i.e.} the transition probability),
	\item $\delta_{s}$ the probability of being in states $e_s$ at $t=1$.
\end{itemize}

To compute the probability of history $\omega_i$, define 
\begin{equation}\label{eq:forward1} 
	\pi_i^{(1)}(s) = 
	  \renewcommand{\arraystretch}{1}
	  \left\{\begin{array}{l @{\quad} l r l}
	        \delta_{s} (1-p_{s,1})  &  \text{if}  &  \omega_{i,1} = 0 \\
	        \delta_{s} (p_{s,1})    &  \text{if}  &  \omega_{i,1} = s
	  \end{array}\right.
\end{equation}

Then for $t = 1, ..., T-1$,
\begin{equation}\label{eq:forwardt}
	\pi_i^{(t+1)}(s) = 
	  \renewcommand{\arraystretch}{1.5}
	  \left\{\begin{array}{l c l}
	        \left[ \sum_{r=1}^S \pi_i^{(t)}(r) \psi_{r,s} \right] (1-p_{s,t+1})  & 
	                        \text{if} & \quad \omega_{i,t+1} = 0 \\
	        \left[ \sum_{r=1}^S \pi_i^{(t)}(r) \psi_{r,s} \right] p_{s,t+1}      &
	                        \text{if} &  \quad \omega_{i,t+1} = s
	  \end{array}\right.
\end{equation}

Note that $\sum_{s=1}^S\pi_i^{(t)}(s)$ is the probability of the history $\omega_i$ until time \emph{t}. Then, the likelihood of history $i$ is 
\begin{equation}\label{eq:pHistMs}
\pi_i= \sum_{s=1}^S\pi_i^{(T)}(s) \, .
\end{equation}

As for model $M_t$, conditioned on the population size, the vector \textbf{y} follows a multinomial with cell probabilities $\pi_i$. \\

\section*{\centering Closed multistate latent model}

The likelihood given by equation \ref{eq:Likelihood} is still valid. So to extend the LMM to the multistate observations, we can modify each part of equation \ref{eq:Likelihood} independently. \\

For the detection part, the likelihood is computed with equation \ref{eq:Likelihoodz}. The probabilities $\pi_k$ are calculated using equations \ref{eq:forward1} to \ref{eq:pHistMs} replacing observed histories $\omega$ by the latent capture histories $\xi$. \\

To account for possible misidentifications, latent error histories $\nu_j$ have to include other values to denotes misidentifications on the different stages. They now include $2S+1$ different values (0 for the unseen, \emph{S} values for the \emph{S} seen states and \emph{S} values for misidentifications on the \emph{S} states). There are  $(2S+1)^T$ latent error histories. The likelihood of the identification process is computed with equation \ref{eq:likelihoodx}, rewriting $A_{j,t} = \alpha^{I(\nu_{j,t} \in [1,S])}  (1-\alpha)^{I(\nu_{j,t} > S)}$. \\

The mechanism to sample $x^{prop}$ stays the same by extending the notations of $\nu_{1t}$, $X_{0,t}$ and $X_{2,t}$ to states. We note $\nu_{1,s,t}$ the history with a unique capture at time $t$ in state $e_s$, $X_{0,s,t}(x)=\{\nu |\nu_t=0, x_\nu>0, x_{\nu_{1,s,t}}>0 \}$ and $X_{2,s,t}(x)=\{\nu |\nu_t=s+S, x_\nu>0\}$. The algorithm from section \emph{Bayesian estimation of the parameters} stays the same with a supplementary first step. This first step is sampling a state $e_s$. Then all following steps from the algorithm are the same, although with the redefined $X_{0,s,t}(x)$ and $X_{2,s,t}(x)$ and for the sampled $s$. \\

\section*{\centering Simulations and analysis}
\subsection*{\centering Simulation design for single state model}

\cite{link_uncovering_2010} have shown that the LMM was effective on one simulation with 5 capture occasions, $\alpha=0.9$ and $\textbf{p}=(0.3,0.4,0.5,0.6,0.7)$ over a population of 400 individuals. To design experiments on large populations or elusive species when capture rates are expected to be low, it seems necessary to know how the model would perform under scenarios with low capture rates and relatively short sequences.
We simulated observation data for $T = 5, 7, 9$, $N = 500, 1000$, $\alpha = 0.8, 0.9, 0.95$ and $p = 0.1, 0.2, 0.3, 0.4$. It makes 24 parameters combinations for each of the three different number of occasions. For the sake of simplicity, we considered the time-dependent model $M_{t,\alpha}$, even though the capture rate was held constant over time in the simulations. \\

\begin{figure}[h]
  	\captionsetup{justification=centering}
	\caption{Beta densities for biased priors on identification probability for the three values used in simulations. The dashed line represents the 95th percentile, the black line the median of the prior and the dotted line the true value of the simulation.}
	\centering
	\includegraphics[]{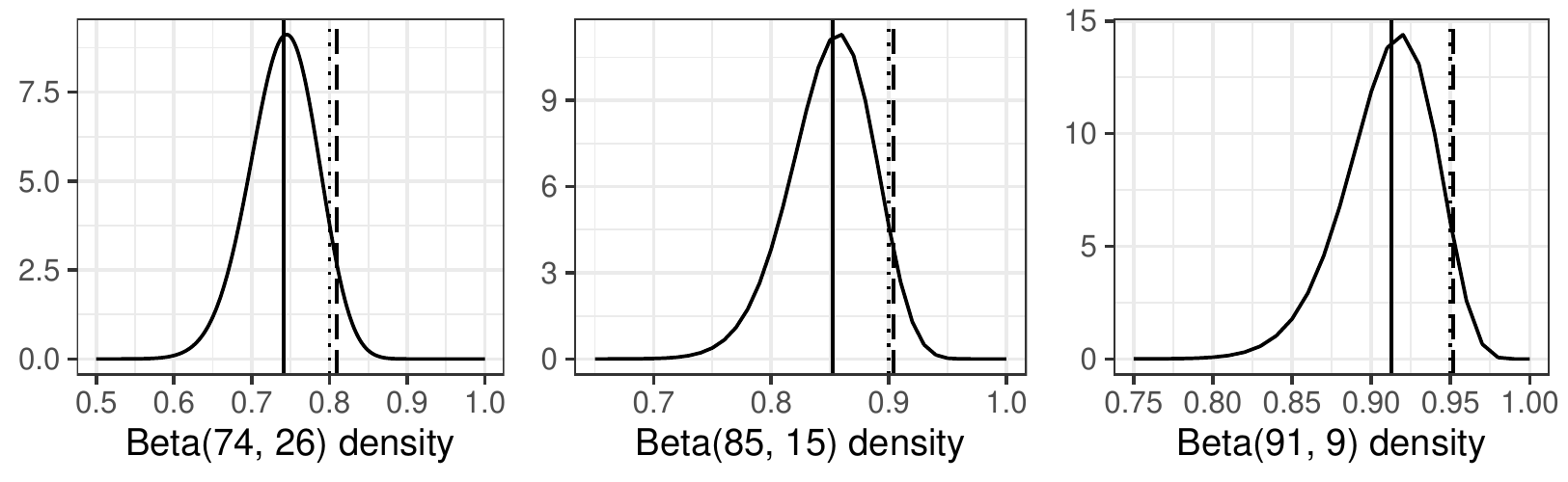}
	\label{betaBiased}
\end{figure}

We expected the model to be weakly identifiable for simulations with low capture rate, making the posterior density unidentifiable. \cite{garrett_latent_2000} define weak identification as the situation where the technical conditions for identifiability are met but the data provides little information about the particular parameters so that their posterior and prior distributions are similar. 
\cite{cole_parameter_2016} say that using informative priors can result in an identifiable posterior when the model is weakly identifiable. We ran the model using three different priors for parameter $\alpha$. 
The first is a non-informative Beta prior.
The other two are informative such as might have been obtained through an evaluation of the identification protocol. Assume that the protocol is run on \emph{n} known individuals and results in $n_a$ correct identifications and $n_b$ errors. The prior is then $\alpha \sim \beta(n_a, n_b)$. We used $n=100$ because it is a convenient value to use and  it is very close to the capacity of a 96-well PCR plate.
The first informative prior was unbiased: for $\alpha$ simulated at $0.8$, we have $\alpha \sim \beta(80, 20)$, for $\alpha$ simulated at $0.9$, $\alpha \sim \beta(90, 10)$ and for $\alpha$ simulated at $0.95$, $\alpha \sim \beta(95, 5)$. 
The second informative prior is a biased version of the first one that underestimates $\alpha$. It is represented on figure \ref{betaBiased}. We chose to underestimate alpha because the model has a tendency to do as such when the capture rate gets too low. The values of $n_a$ and $n_b$ were chosen such as the true value used for the simulation lies around the 95th percentile of the prior distribution (dashed line on figure \ref{betaBiased}). They are as following: $\alpha_{simulated(0.8)} \sim \beta(74, 26)$, $\alpha_{simulated(0.9)} \sim \beta(85, 15)$ and $\alpha_{simulated(0.95)} \sim \beta(91, 9)$. These priors have respective means of 0.74, 0.85 and 0.91.
In order to study the effect of the prior on $\alpha$ over the model, we calculated the overlap $\tau$ between this prior and the estimated posterior as suggested by \cite{garrett_latent_2000}. \\

\subsection*{\centering Simulation design for multistate model}

For the multistate model, the same design as for single state was used. We considered three states with possibility of transition between all states. For the sake of comparison, the transition matrix used is taken from \parencite{worthington_estimation_2019} as:
\[
\boldsymbol\phi=
\begin{pmatrix}
0.76 & 0.12 & 0.12\\
0.1 & 0.8 & 0.1\\
0.15 & 0.15 & 0.7
\end{pmatrix}
\]
and the initial states are fixed to its equilibrium distribution, that is $\boldsymbol\delta = (0.33, 0.4, 0.27)$. \\

\section*{\centering Implementation}

We used NIMBLE \parencite{de_valpine_programming_2017} to implement the model. Unlike Jags (for example), NIMBLE allows new distributions as well as all samplers for the MCMC to be written as we need. We needed it to code the likelihood of the model and to code the sampler of \textbf{x}. We were also able to write all the Gibbs samplers previously detailed for a maximum computational efficiency. 
In order to improve efficiency, all observable histories which had zero count were not considered \emph{i.e.} their corresponding rows and columns in matrices \textbf{A} and \textbf{B} were deleted as suggested in \cite{schofield_connecting_2015}.
For the single state simulations, the MCMC was run over 1E6 iterations after a burn-in period of 20,000 iterations (30,000 for $\alpha = 0.8$) and the chains were thinned by a factor of 1/200 in order to limit memory usage. 
For the multistate simulations, the computational cost per iteration is much higher so we only ran 500,000 iteration with a thinning of 1/100 and an additional burnin of 60,000 iterations. For most simulation scenarios, this proves to be enough. For simulations where $T=5$ as well as where $T=7$, $p \le 0.2$, we instead had to use more iterations. We used 1E6 iteration with a thinning of 1/200 and an additional burnin of 100,000 iterations.
We ran two chains for each simulation with two different starting points. For the first one, \textbf{x} was initialized as the set of observed histories, as if there was no error. In the second one, we arbitrarily added 40 errors randomly.
For dealing with the unobserved individuals, we wrote the likelihood conditional on the population size. We added the parameter $n_0$ to denote the number of unseen individuals. \\

\section*{\centering Results}
We checked the convergence with Rhat and the visual of the chains of the parameter \emph{N} (the slowest to converge and the one with the highest autocorrelation).

\subsection*{\centering Single state model results}

\begin{figure}[h]
  	\captionsetup{justification=centering}
	\caption{Single state population size estimations (y axis) depending on capture probability (x axis), identification probability, number of capture occasion (on the left) and prior on the identification probability (on top). Horizontal dashed lines indicate true population size. Grey points are simulation mean-estimates. Black points are averaged estimates. Empty triangles are the averaged estimate of the 97.5\% and 2.5\% quantiles.}
	\centering
	\includegraphics{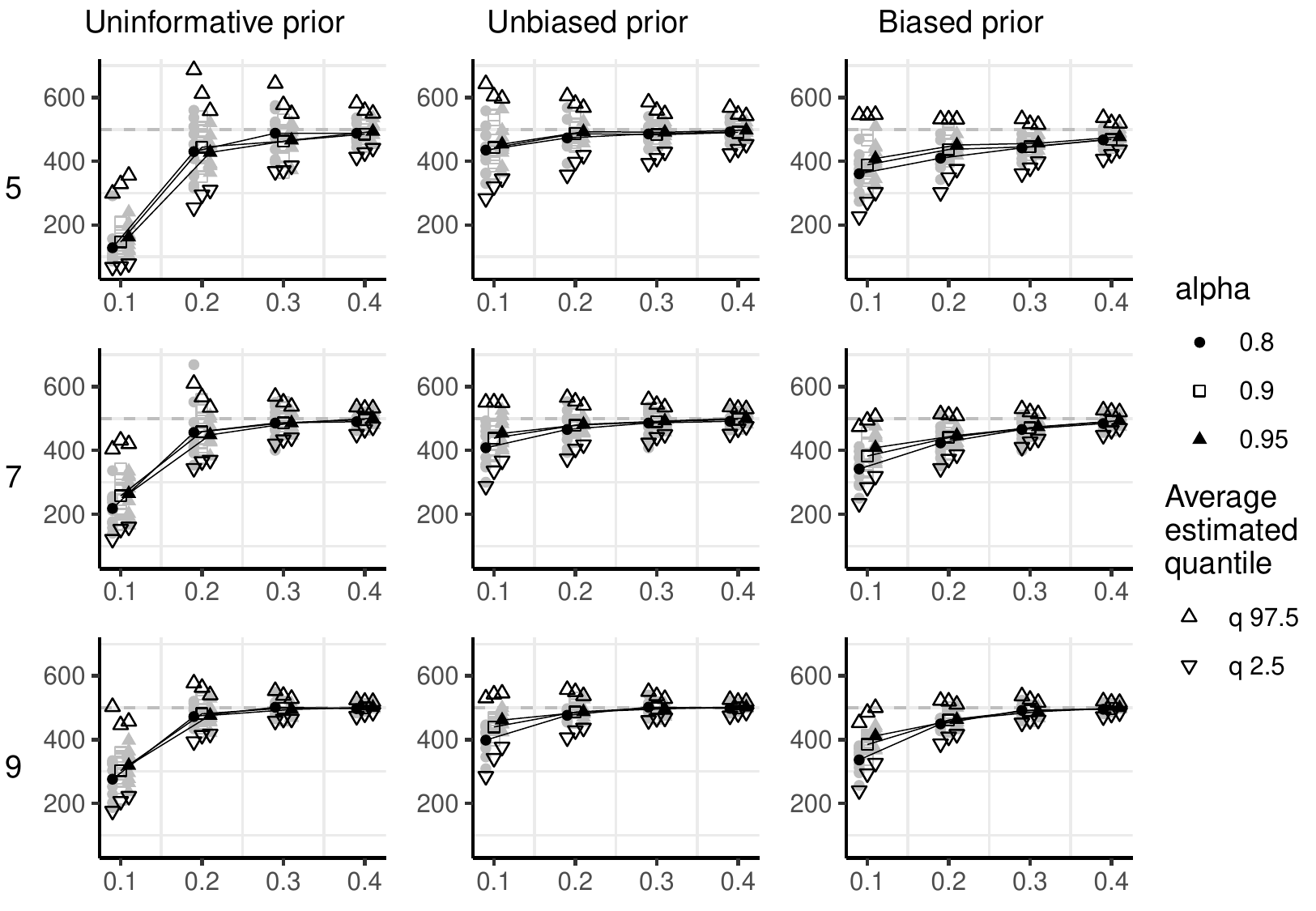}
	\label{EstimN}
\end{figure}

\begin{figure}
  	\captionsetup{justification=centering}
	\caption{Boxplots of the overlap value beween the prior and posterior of the identification probability. The horizontal line is at 0.35 (see  \cite{garrett_latent_2000}). On the x-axis legend, the letter 'a' stands for the identification probability and the letter 'p' for the capture probability, the corresponding values simulated following.}
	\centering
	\includegraphics[]{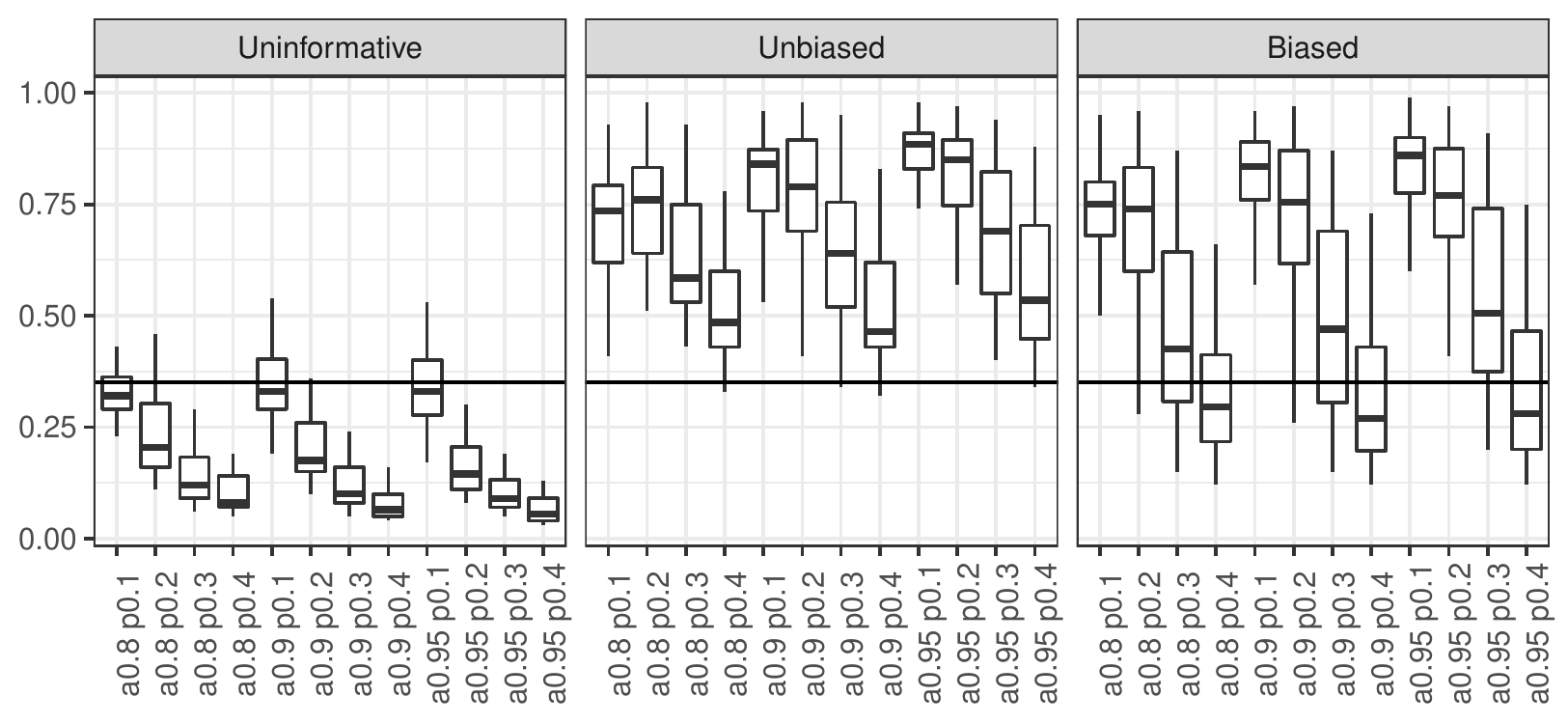}
	\label{overlaps}
\end{figure}

The population size estimation with a single state and $N=500$ is shown in figure \ref{EstimN}. Using the uninformative prior, no bias was observed for $p \ge 0.3$. When $p=0.2$, the average relative bias goes from 3\% (when T=9) to 14\% (when T=5). When convergence was reached for simulations with $p=0.1$, the average relative bias was over 30\% when T=9 and over 40\% when T=7.
When adding the unbiased prior, for $p=0.1$, the population size is underestimated by about 10\% on average but this bias rises to 40\% for some simulations. Also, for 80\% of the simulations with $p=0.1$, the real population size lies in the estimated 95\% interval.
The use of the biased prior does not affect the estimations for $p=0.4$. But as \emph{p} decreases, the population size gets more underestimated. The average bias goes down to 32\% for the lowest values of \emph{p} and $\alpha$ with 9 capture sessions. 
Higher values of $\alpha$ lead to a reduced bias when it occurs and a reduced confidence interval.
The results are very similar for $N = 1000$ only slightly better. \\

When looking at the overlaps between a prior and a posterior, \cite{garrett_latent_2000} give the value of 0.35 as a guide, over which a model is weakly identified. 
We show the overlaps between prior and posterior of $\alpha$ in figure \ref{overlaps}. With the uninformative prior, all simulations with $p \ge 0.3$ and most of the ones with $p=0.2$ result in an overlap between prior and posterior for $\alpha$ that is lower than 0.35.
With the informative priors, for most of the simulations, the prior and posterior of $\alpha$ are highly overlapping and almost confounded for low recaptures. 
The informative priors overlap less with their corresponding posterior for $p \ge 0.3$. \\

Running two chains with 1,030,000 iterations on a 3.0GHz Intel processor took less than five minutes, even with $T=9$. 
With the uninformative prior on $\alpha$, convergence was achieved for all simulations with a capture rate of 0.3 or above. For $T=5$, with $p=0.2$ some chains did not converged while with $p=0.1$ none did. Increasing $T$ to 7 did allow for a better convergence with $p=0.2$ but not with $p=0.1$. Finally, $T=9$ resulted in good convergence for more than half the simulations with $p=0.1$. In addition, convergence was slower for lower values of $\alpha$ and for $N=1000$.
There is a high autocorrelation for the N-chains that makes some of them have an effective sampling size less than 100.
When an informative prior on $\alpha$ is used, the chains always converge and the effective sampling size is always over 75 (average is over 200). \\

\subsection*{\centering Multistate model results}

\begin{figure}[h]
  	\captionsetup{justification=centering}
	\caption{Multistate population size estimations (y axis) depending on capture probability (x axis), identification probability (point shape), number of capture occasions (on the left) and identification probability prior (on top). Grey points are simulation mean-estimates. Black points are averaged estimates. Horizontal dashed lines indicate true population size.}
	\centering
	\includegraphics[]{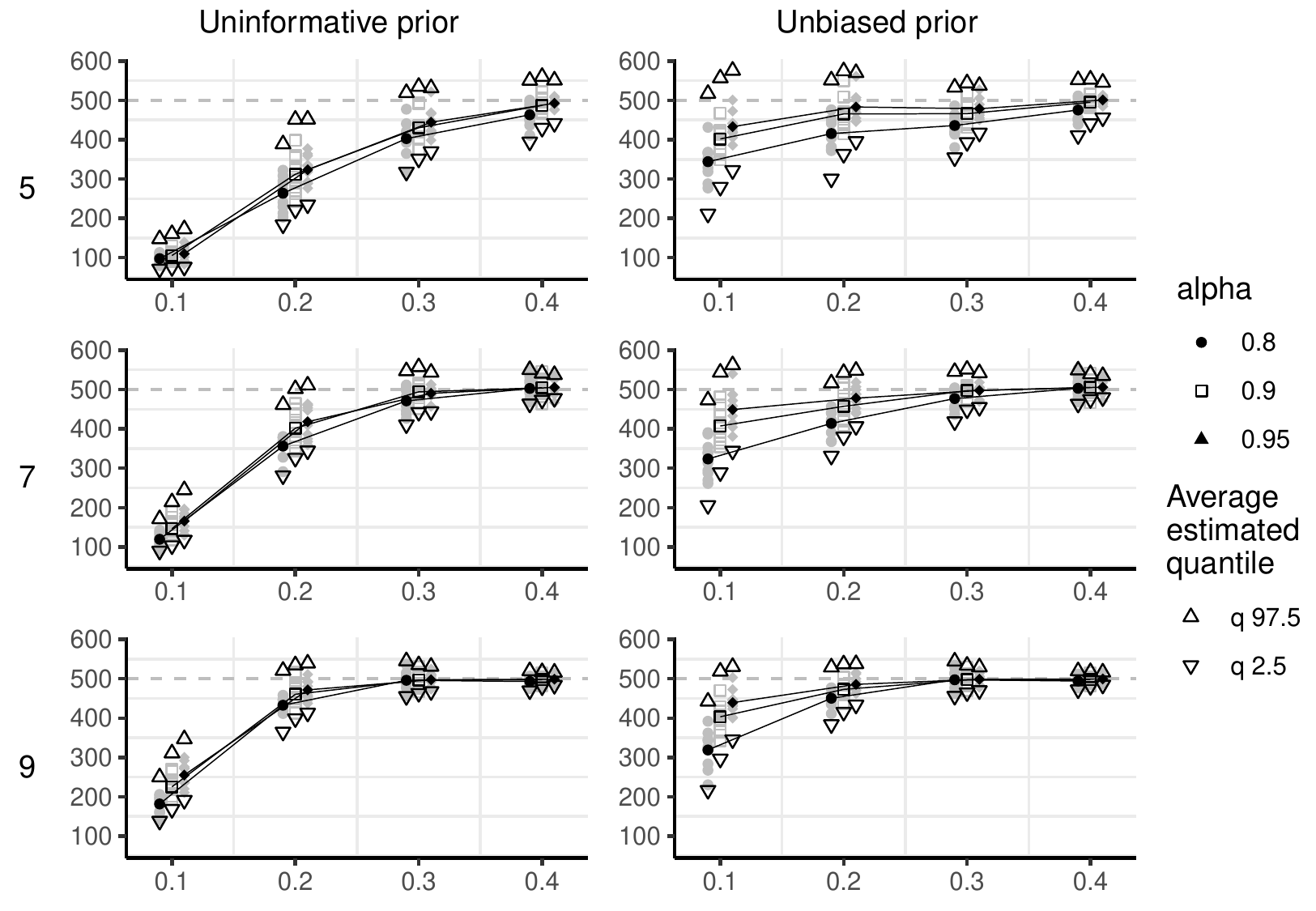}
	\label{EstimMulti}
\end{figure}

Multistate population size estimation for $N=500$ are shown on figure \ref{EstimMulti}. Using the uninformative prior, no bias is observed for $p \ge 0.4$.
For $T=5$, the estimates are biased as soon as $p \le 0.3$. The average relative bias ranges from 10\% (for $p=0.3, \, \alpha=0.95$) to 50\% (for $p=0.2, \, \alpha=0.8$).
When $T=7$, the estimates are slightly biased (5\% at most) for $p=0.3$. Results show more bias for lower capture rates, bias ranging between 16\% and 30\% for $p=0.2$.
When $T=9$, the estimates are biased only for $p\le0.2$, bias ranging between 9\% and 14\%.
When adding the unbiased prior, the average relative bias is reduced. For $p=0.2, \, \alpha=0.8$, it is reduced to 10\% for $T=9$ and to 17\% for $T=7$ and $T=5$. 
The results for $N = 1000$ are similar although the bias is reduced. \\

The estimations of transitions probabilities are globally unbiased for $p \ge 0.3$  or for $T = 9$. Some transitions have an average bias that is always under 0.1. The relative bias can be quite high for low probability transition but the estimation always lies in the 95\% interval. For $p = 0.2$ the size of this interval is around 0.4, the estimates are thus very imprecise. Finally adding an informative prior on $\alpha$ does not change the estimates of the transitions probabilities nor the size of the estimated intervals. \\

Running two chains of 1,100,000 iterations, on the same processor as for single-state, took around 4 hours for $T=9$.
With the uninformative prior on $\alpha$, convergence was achieved for all chains except the ones where $T=5$ and $p=0.2$. For these ones, adding an informative prior led to proper convergence. Some more iterations are needed for $N=1000$ as a lot of chains have an effective sampling size under 100. \\

\section*{\centering Discussion}

We conducted a simulation analysis to help design CMR experiment where eDNA is to be used for identification and where low capture rates are anticipated. We showed, in single and multi state experiments, on which range of parameters the LMM could be safely used for population size and transition rates estimations in a closed population. 
When the capture rates and the number of capture occasions are too low, the model is weakly identified. \cite{carlin_bayes_1996} says that, in this case, there is a high cross-correlation that leads to very slow convergence. When the probability of identification $\alpha$ decreases, this problem is amplified and additionally, the estimates are less precise.
This demonstrates that using the LMM does not solve completely the problem of identification but should be used in parallel with experimental reduction of the errors. 
Although the use of an informative prior does not guarantee the identifiability of a weakly identified model, it appears to be the case for our simulations since convergence is always reached when using one, even a biased one. Considering that the priors we used were strongly informative, in cases with low recaptures were the data does not inform on $\alpha$, it may seem reasonable to remove the parameter by fixing its value, rather than trying to estimate it. Finally, sensibility to this prior should be tested since with enough captures the biased prior lead to estimates slightly biased compared to using an uninformative one. \\

In this paper, we implemented the LMM of \cite{link_uncovering_2010} using Nimble and the sampling algorithm of \cite{bonner_extending_2015}. This allowed for much faster MCMC than what \cite{link_uncovering_2010} reported. This work is a first step toward the accessibility of the model at a larger scale. The nimble implementation will allow others to use the model for their own needs. However, for cases different from what has been presented here, some changes must be done, both in the model code and in the samplers code. We intend to make the codes functional for broader situations as well as easier to use through a package. \\

It is necessary to keep in mind that the model assumes that ghosts can only be generated once and thus cannot be resighted. This hypothesis might not hold in some cases, making the model as it is not usable for them. 
It is also useful to note that the framework of the LMM is not limited to closed population and can be modified to estimate survival. This is accomplished by replacing the likelihood of the capture process [\textbf{z} | N, \textbf{p}] by the likelihood of an open population model, such as the Cormack-Jolly-Seber model (CJS) [\textbf{z} | $\phi$, \textbf{p}]. \cite{bonner_extending_2015}, developed such a model with a different kind of misidentification (an individual is misidentified as an other one that has been seen at least once before) and we are currently working on a multistate open population with misidentifications such as in this paper. 
The model can also be extended with data augmentation in order to account for capture heterogeneity between individuals as in \cite{mcclintock_probit_2014}.  Additionally, we are working on using additional information about the data used for identifications. In particular we are trying to use the quality of a sample as a covariate of identification. \\

For studies using eDNA in order to identify individuals, this paper shows that more samples could be kept or even collected. The LMM makes it possible to allow for about 5 to 10\% of misidentifications and have good estimates of the parameters. Low capture rate can be compensated for if prior information about the misidentifications is available.
The LMM is especially promising for studying large populations or very elusive species since increasing the capture effort could then be expensive compared to keeping samples.
Additionally, there is potential  for new experiments where lower quality samples would be obtained, provided eDNA can be sampled. An example of such a study would be on insects such as mosquitoes, as in the project that motivated this paper. \\

\section*{\centering Acknowledgements}
We thank Daniel B. Turek at Williams College Department of Mathematics and Statistics for his quick and efficient help with nimble.

\section*{\centering Conflict of interest disclosure}
The authors declare that they comply with the PCI rule of having no financial conflicts of interest in relation to the content of the article. 

\section*{\centering Data, script, code, and supplementary information availability}
Data, script and codes are available online: DOI of the webpage hosting the data \url{https://doi.org/10.5281/zenodo.7794259}.

\titleformat*{\section}{\bfseries\Large\centering}

\printbibliography[notcategory=ignore]

\end{document}